\documentclass[aps,preprint,amssymb]{revtex4}

\usepackage{epsfig}
\usepackage{bm}
\begin{document}
\renewcommand{\thefootnote}{\fnsymbol{footnote}} 
\renewcommand{\theequation}{\arabic{section}.\arabic{equation}}

\title{Time-dependent current density functional theory via time-dependent deformation functional theory: A constrained search formulation in the time domain}

\author{I. V. Tokatly}
\email[]{E-mail: ilya_tokatly@ehu.es}

\affiliation{European Theoretical Spectroscopy Facility (ETSF),\\
Departamento de Fisica de Materiales, Universidad del Pais Vasco UPV/EHU, San Sebastian, Spain, \\
and Moscow Institute of Electronic Technology, Zelenograd, Moscow, Russia}

\date{\today}

\begin{abstract}
\noindent The logical structure and the basic theorems of time-dependent current density functional theory (TDCDFT) are analyzed and reconsidered from the point of view of recently proposed time-dependent deformation functional theory (TDDefFT). It is shown that the formalism of TDDefFT allows to avoid a traditional external potential-to-density/current mapping. Instead the theory is formulated in a form similar to the constrained search procedure in the ground state DFT. Within this formulation of TDCDFT all basic functionals appear from the solution of a constrained universal many-body problem in a comoving reference frame, which is equivalent to finding a conditional extremum of a certain universal action functional. As a result the physical origin of the universal functionals entering the theory, as well as their proper causal structure becomes obvious. In particular, this leaves no room for any doubt concerning predictive power of the theory.
\end{abstract}

\maketitle

\section{Introduction}
\label{intro}

Last decades witness a growing popularity of the time-dependent density functional theory (TDDFT) \cite{RunGro1984} as a practical tool for studying dynamics of various quantum many-body systems (see \cite{TDDFT2006} and references therein). The range of applications of TDDFT (including its current-based version, the time-dependent current density functional theory (TDCDFT), \cite{GhoDha1988,Vignale2004}) is impressively broad. Interaction of atoms, molecules and solids with electro-magnetic radiation, transport phenomena in nanostructured systems, dynamics of nuclear matter and ultracold trapped atomic gases form by far not a complete list of modern TDDFT/TDCDFT applications. 

The first fundamental contribution to TDDFT, which can be also considered as a birth of the theory, was made by Runge and Gross in 1984 \cite{RunGro1984}. Many importand developments and generalizations of TDDFT which appeared in the last 25 years deepened significantly our understanding of this approach (a good discussion of the current state of a formal theory can be found in a review \cite{vanLeeuwen2001} and in Chapters~1-7 of the book \cite{TDDFT2006}). Nonetheless surprising observations, fundamental puzzles as well as misunderstandings related to formal aspects of TDDFT regularly appear up to now. The most known and, possibly, historically the first puzzle of this kind was a so called causality paradox closely related to problems with variational definition of xc potentials in TDDFT. Presently we have many various explanations and resolutions of this paradox \cite{Rajagopal1996,vanLeeuwen1998,vanLeeuwen2001,Mukamel2005,TDDFT2006}, but, surprisingly, the most straightforward and elementary resolution of the problem appeared only in the last year \cite{Vignale2008}. Another puzzling surprise that appeared in the last year was and observation made by Baer \cite{Baer2008}. He has found that in a lattice formulation of the theory, in contrasts to the ground state DFT, there are apparently reasonable time-dependent densities which are not v-representable. This problem was discussed and clarified in two very resent papers \cite{LiUll2008,Verdozzi2008}, but it became quite clear that v-representability, which is commonly taken for granted in TDDFT and TDCDFT, can be a serious issue. The second fundamental issue of TDDFT, the density-to-potential mapping, was also under debate in the last year \cite{SchDre2007,MaiLeeBur2008}. The authors of Ref.~\cite{SchDre2007} attempted to analyze the causal structure of the mapping from the density to external potential and arrived to the conclusion that TDDFT in the Kohn-Sham (KS) formulation can not predict the dynamics of the density as the xc potential does depend on a ``future'' in a form of a second time derivative of the density. In a recent comment Maitra, et.~al. \cite{MaiLeeBur2008} have demonstrated that this conclusion is based on a misunderstanding of the basic mapping theorem \cite{vanLeeuwen1999}, and proved that KS-TDDFT has a full predictive power. However at the present stage of the theory this can be shown only for the KS-TDDFT, while for the other, for example, hydrodynamic formulations \cite{RunGro1984,TokatlyPRB2005b} of the theory the question raised in Ref.~\cite{SchDre2007} is still formally valid. In general we should honestly say that the causal structure of all basic functionals in TDDFT and TDCDFT is not well studied and understood. Hence in the last year we have seen the whole bunch of papers debating actually all most fundamental points of TDDFT and related approaches. Therefore it is timely to carefully reexamine the foundations and the logical structure of the theory and formulate it, from the very beginning, in a way that takes into account our current level of understanding and the most recent developments in the field.

This work is an attempt of such reexamination. Below I concentrate of the current density version of the theory, having in mind its generality and numerous recent applications of TDDFT-based methods to the transport theory where the exact knowledge of the current is vital. From conceptual point of view TDCDFT is a reduced theory which allows one to describe the behavior of the current ${\bf j}$ and the density $n$ formally ignoring the rest of complicated microscopic dynamics of a quantum many-body system. In other words TDCDFT is a closed theory of convective motion. Therefore it is most natural to formulate such a theory in a form of a closed system of equations of motion for the basic variables, the density and the current. This is the hydrodynamic formulation of TDCDFT, which analyzed in detail in Sec.~\ref{TDCDFT-hyd}. I show how hydrodynamic equations of motion are related to the microscopic many-body dynamics, and how one can make them closed by employing the traditional formulations of the TDCDFT mapping theorem \cite{GhoDha1988,Vignale2004}. The main concern of Sec.~\ref{TDCDFT-hyd} is to prove an equivalence of using different sets of basic variables which can be used in TDCDFT to describe the convective degrees of freedom. These are the current and the density, $({\bf j},n)$, the velocity field and the density, $({\bf v},n)$, in the Eulerian description of convective dynamics, or a continuous set of trajectories ${\bf x}({\bm\xi},t)$ of infinitesimal fluid element in the Lagrangian description. The hydrodynamic equations of TDCDFT for all three possible choices of basic variables are derived and compared. I also derive a KS formulation of TDCDFT and show how the basic universal functionals entering the hydrodynamic TDCDFT and KS-TDCDFT are related to each other. The most important part of the paper is Sec.~\ref{TDDefFT} where the foundations of TDCDFT are reconsidered from the point of view of recently proposed time-dependent deformation functional theory (TDDefFT) \cite{TokatlyPRB2007}. I prove the equivalence of two theories and present a new logical structure of TDCDFT which is based on TDDefFT. The idea is very simple and physically transparent. From the very beginning we take the Lagrangian trajectories as a set of basic variables for a future closed theory of the convective motion, and separate the convective degrees of freedom by transforming the many-body theory to a reference frame moving with fluid elements. After this transformation an appearance of a closed theory of the convective motion becomes obvious. In fact, TDDefFT and thus TDCDFT take a form which very similar to the constrained search formulation of the ground state DFT \cite{Levy1979,Levy1982,Lieb1983}.

In this paper I am trying to avoid long algebraic manipulations and simply state the results when they are physically plausible (some detailed calculations, especially those related formalities of nonlinear transformations of coordinates can be found in \cite{TokatlyPRB2005a,TokatlyPRB2005b,TokatlyPRB2007}. Instead, I am concentrating on the logical structure of the theory and on discussions of some delicate issues, such as equivalence of different descriptions, uniqueness and existence of solutions, etc. I hope the this way of presentation is more suited to the purpose of this work.

\section{Hydrodynamic formulation of TDCDFT}
\label{TDCDFT-hyd}

\subsection{Starting point: General formulation of the quantum many-body problem}

Let us consider a system of $N$ identical particles in the presence of time dependent external scalar $U({\bf x},t)$ and vector ${\bf A}({\bf x},t)$ potentials. The corresponding many-body wave function 
$\Psi({\bf x}_{1},\dots,{\bf x}_{N},t)$
is a solution to the time-dependent Schr\"odinger equation
\begin{equation}
 i\partial_t\Psi({\bf x}_{1},\dots,{\bf x}_{N},t) = H\Psi({\bf x}_{1},\dots,{\bf x}_{N},t)
\label{SE}
\end{equation}
with the following Hamiltonian 
\begin{equation} 
H = \sum_{j = 1}^{N}\left[\frac{(i\partial_{{\bf x}_{j}} 
+ {\bf A}({\bf x}_{j},t))^{2}}{2m} + U({\bf x}_{j},t)\right] +
\frac{1}{2}\sum_{j\ne k} V(|{\bf x}_{j}-{\bf x}_{k}|)
\label{H}
\end{equation}
where $V(|{\bf x}-{\bf x}'|)$ is the interaction potential. For a given initial condition, 
\begin{equation}
\Psi({\bf x}_{1},\dots,{\bf x}_{N},0) = \Psi_0({\bf x}_{1},\dots,{\bf x}_{N}),
 \label{InitialPsi}
\end{equation}
the dynamics of the system is completely specified by Eq.~(\ref{SE}).

In most physically important situations it is not necessarily to know the full many-body wave function. Normally the experimentally measurable response of the system to external probes can be described in terms reduced ``collective'' variables -- the density of particles $n({\bf x},t)$, and the density of current ${\bf j}({\bf x},t)$
\begin{eqnarray}
 \label{n}
n({\bf x},t) &=& \rho({\bf x},{\bf x},t),\\
\label{j}
{\bf j}({\bf x},t) &=& \frac{i}{2m}\lim_{{\bf x}'\to{\bf x}}(\partial_{\bf x} -\partial_{\bf x'})\rho({\bf x},{\bf x}',t) 
-\frac{n}{m}{\bf A}({\bf x},t),
\end{eqnarray}
where $\rho({\bf x},{\bf x}',t)$ is the one particle reduced density matrix
\begin{equation}
 \label{rho}
\rho({\bf x},{\bf x}',t) = N \int \prod_{j =2}^{N}d{\bf x}_{j}
\Psi^*({\bf x},{\bf x}_{2},\dots,{\bf x}_{N},t)\Psi({\bf x}',{\bf x}_{2},\dots,{\bf x}_{N},t)
\end{equation}

The main idea of TDCDFT is to reduce, at the formally exact level, the problem of calculation of the density and the current to solving a closed system of equations which involve only $n({\bf x},t)$ and ${\bf j}({\bf x},t)$. In the next two subsections we describe a few equivalent ways to formulate such a closed theory.

\subsection{Local conservation laws and TDCDFT hydrodynamics in Eulerian formulation}
\label{Euler-hyd}

Using the microscopic definitions of Eqs.~(\ref{n}) and (\ref{j}), and the Schr\"odinger equation (\ref{SE}) one can derive the following hydrodynamic equations of motion the density and the current
\begin{eqnarray}
 \label{continuity1}
&&\partial_t n + \partial_{x^{\mu}}j_{\mu}= 0,\\
\label{NS1}
&&m\partial_t j_{\mu} - [{\bf j}\times{\bf B}]_{\mu} - nE_{\mu} + \partial_{x^{\nu}}\Pi_{\mu\nu}=0
\end{eqnarray}
where ${\bf E}({\bf x},t)$ and ${\bf B}({\bf x},t)$ are electric and magnetic fields generated by the external time-dependent scalar and vector potentials
\begin{eqnarray}
{\bf E}({\bf x},t) &=& - \partial_{t}{\bf A}({\bf x},t) 
- \partial_{\bf x}U({\bf x},t),
\label{E}\\
{\bf B}({\bf x},t) &=& \partial_{\bf x}\times{\bf A}({\bf x},t).
\label{B}
\end{eqnarray}
Equation (\ref{continuity1}) is the usual continuity equation, i.~e., a local conservation law of the number of particles. The equation of motion for the current, Eq.~(\ref{NS1}), physically corresponds to a local  momentum conservation law (or a local force balance): the time derivative of the current equals to a sum of the external and internal forces. Importantly, the local internal force (the last term in Eq.~(\ref{NS1})) has a form of a divergence of a second rank tensor. Therefore the bulk internal force vanishes after a volume integration, as it is required by the Newton third law.
The momentum flow tensor $\Pi_{\mu\nu}$ entering Eq.~(\ref{NS1}) contains a kinetic and an interaction contributions \cite{PufGil1968,TokatlyPRB2005a}, $\Pi_{\mu\nu}= \Pi_{\mu\nu}^{\rm kin}+\Pi_{\mu\nu}^{\rm int}$, which are expressed in terms of the many-body wave function as follows
\begin{eqnarray}
 \label{Pi_kin}
&& \Pi_{\mu\nu}^{\rm kin}({\bf x},t) = \frac{1}{2m}\left[\lim_{{\bf x}'\to{\bf x}}
(\hat{P}^*_{\mu}\hat{P}'_{\nu}+\hat{P}^*_{\nu}\hat{P}'_{\mu})\rho({\bf x},{\bf x}',t) 
- \frac{\delta_{\mu\nu}}{2}\partial_{\bf x}^2n({\bf x},t)\right],\\
\label{Pi_int}
&& \Pi_{\mu\nu}^{\rm int}({\bf x},t) = - \frac{1}{2}\int d{\bf x}' \frac{x'^{\mu}x'^{\nu}}{|{\bf x}'|}
\frac{\partial V(|{\bf x}'|)}{\partial |{\bf x}'|}\int_{0}^{1} d\lambda
\Gamma({\bf x}+\lambda{\bf x}',{\bf x}-(1-\lambda){\bf x},t),
\end{eqnarray}
where $\hat{P}_{\mu} = -i\partial_{x^{\mu}} - A_{\mu}({\bf x},t)$ is the kinematic momentum operator, and $\Gamma({\bf x},{\bf x}',t)$
is a two particle reduced density matrix
\begin{equation}
 \label{Gamma}
\Gamma({\bf x},{\bf x}',t) = N(N-1)\int \prod_{j =3}^{N}d{\bf x}_{j}
\Psi^*({\bf x},{\bf x}',{\bf x}_{3},\dots,{\bf x}_{N},t)\Psi({\bf x},{\bf x}',{\bf x}_{3},\dots,{\bf x}_{N},t).
\end{equation}

A formal possibility to formulate a closed theory for calculation the density and current distributions follows from the mapping theorem of TDCDFT \cite{GhoDha1988,Vignale2004}. The main statement of this theorem can be formulated as follows. For a given initial state $\Psi_0$ a map of the external potentials to the current, $(U,{\bf A})\mapsto{\bf j}$, is invertible and unique up to a gauge transformation, provided the potentials are analytic in time, and the current density is v-representable. (Note that here the term ``v-representability'' is understood in a broad sense: a current ${\bf j}$ is called v-representable if it can be produced by some external 4-potential $(U,{\bf A})$). 

In fact, the above mapping theorem states that the potentials are unique (modulo a gauge transformation) functionals of the initial state and the current density, $U = U[\Psi_0,{\bf j}]$ and ${\bf A} = {\bf A}[\Psi_0,{\bf j}]$. This immediately implies that the many-body wave function $\Psi(t)$ and, therefore, any physical observable is also a functional of $\Psi_0$ and ${\bf j}$. In particular, inserting the functionals ${\bf A}[\Psi_0,{\bf j}]$ and $\Psi = \Psi[\Psi_0,{\bf j}]$ into the definitions of Eqs.~(\ref{Pi_kin}) and (\ref{Pi_int}) we get the exact momentum flow tensor as a unique functional of the initial wave function and the current $\Pi_{\mu\nu}= \Pi_{\mu\nu}[\Psi_0,{\bf j}]$. This functional is universal in a sense that it does not explicitly contain the external potentials, but is uniquely recovered from a given current and an initial state (in the following for the sake brevity we omit $\Psi_0$ in the arguments of the functionals). Substituting the functional $\Pi_{\mu\nu}[{\bf j}]$  into Eq.~(\ref{NS1}) we obtain a closed system of equations of motion for $n({\bf x},t)$ and ${\bf j}({\bf x},t)$. Hence from the system of Eqs.~(\ref{continuity1})-(\ref{NS1}) we can in principle determine the dynamics of the density of particles and the density of current, avoiding, at least formally, the explicit solution of the full many-body problem. TDCDFT represented in a form of the closed system (\ref{continuity1})-(\ref{NS1}) can be viewed as an exact quantum hydrodynamics. It is worth noting that the hydrodynamic formulation of TDCDFT is analogous to the formulation of the static DFT in a form of the direct Hohenberg-Kohn variational principle \cite{HohKohn1964}. 

A connection of the TDCDFT hydrodynamics to the standard mechanics fluids \cite{LandauVI:e} can be made more obvious if we switch the basic variable from the current ${\bf j}$ to the velocity field ${\bf v}={\bf j}/n$. It is also useful to extract from the full momentum flow tensor $\Pi_{\mu\nu}$ its exactly known part -- the flow of momentum due to convective motion of the fluid, $mnv_{\mu}v_{\nu}$,
\begin{equation}
 \label{P_conv}
\Pi_{\mu\nu} = mnv_{\mu}v_{\nu} + P_{\mu\nu}
\end{equation}
where $P_{\mu\nu}$ is the stress tensor which is responsible for a local internal force related to a relative motion of particles ``inside'' an small moving fluid element (see Sec.~\ref{TDDefFT} below for a more detailed discussion). Using the representation (\ref{P_conv}) and expressing all currents in terms of the velocity field we transform equations of motion (\ref{continuity1}) and (\ref{NS1}) to the following standard ``Navier-Stokes'' form
\begin{eqnarray}
 \label{continuity2}
(\partial_t + {\bf v}\partial_{\bf x}) n &+& n\partial_{x^{\mu}}v_{\mu}= 0,\\
\label{NS2}
m(\partial_t + {\bf v}\partial_{\bf x})v_{\mu} &-& [{\bf v}\times{\bf B}]_{\mu} - E_{\mu} + \frac{1}{n}\partial_{x^{\nu}}P_{\mu\nu}[{\bf v}]=0.
\end{eqnarray}
Since the map $(n,{\bf j})\mapsto (n,{\bf v})$ is one-to-one we are allowed to replace the functional dependence of the stress tensor on the current by the functional dependence dependence on the velocity field, $P_{\mu\nu}[{\bf j}]\mapsto P_{\mu\nu}[{\bf v}]$. It is also worth noting the knowledge of only the current ${\bf j}$ or the velocity ${\bf v}$ is sufficient to recover the density by integrating the continuity equation of the form of (\ref{continuity1}) or (\ref{continuity1}), respectively. 

\subsection{Kohn-Sham construction in TDCDFT}

Practical applications of any DFT always rely on the Kohn-Sham (KS) construction \cite{KohnSham1965}, which in the present time-dependent setting can be introduced as follows (see, e.~g., \cite{TokatlyPRB2005b}). Let us consider a fictitious system of $N$ noninteracting particles in the presence of an electromagnetic field generated by the external 4-potential $(U,{\bf A})$, and by some selfconsistent vector ${\bf A}^{\rm xc}$ and scalar $U^{\rm xcH}=U^{\rm xc}+U^{\rm H}$ potentials, where $U^{\rm H}$ is the usual Hartree potential. The dynamics of this system is described by a set of one-particle Schr\"odinger equations for KS orbitals $\phi_j({\bf x},t)$, $j=1\dots N$
\begin{equation}
 \label{KS}
i\partial_t \phi_j = \frac{1}{2m}(i\partial_{\bf x} + {\bf A}+{\bf A}^{\rm xc})^2\phi_j + (U+U^{\rm xcH})\phi_j
\end{equation}
Obviously the density $n_S$ and the velocity ${\bf v}_S$ of the KS system satisfy the continuity equation (\ref{continuity2}) and the force balance equation of the form of Eq.~(\ref{NS2}), but with the stress tensor $P_{\mu\nu}$, and the external Lorentz force being replaced, respectively, by the kinetic stress tensor $T_{\mu\nu}^{S}$ of the noninteracting KS particles, and by the Lorentz force corresponding to the total effective 4-potential. From the requirement that the KS density and current reproduce the density and the current in the real interacting system we get the following selfconsistency equation connecting the xc potential to the stress tensor functional
\begin{equation}
 \label{Axc}
\partial_t A^{\rm xc}_{\mu} - [{\bf v}\times (\partial_{\bf x}\times{\bf A}^{\rm xc})]_{\mu} + \partial_{x^{\mu}}U^{\rm xcH}
= - \frac{1}{n}\partial_{x^{\nu}}\Delta P_{\mu\nu}[{\bf v}],
\end{equation}
where $\Delta P_{\mu\nu}[{\bf v}]=P_{\mu\nu}[{\bf v}] - T^{S}_{\mu\nu}[{\bf v}]$ is a difference of the stress tensors in the interacting and noninteracting KS systems. Equation (\ref{Axc}) can serve as a most general definition of the xc potentials. For a given stress tensor functional (the right hand side) it defines the xc 4-potential $(U^{\rm xc},{\bf A}^{\rm xc})$ up a gauge transformation. 

It is important to stress out that the KS construction is only an auxiliary formal device for solving the general collective variable theory in a form of closed equations of motion, Eqs.~(\ref{continuity1}) and (\ref{NS1}) ( or, equivalently, Eqs.~(\ref{continuity2}) and (\ref{NS2})), for basic variables -- the current/velocity and the density. The situation is very similar to the static DFT where the KS costruction serves merely as useful mathematical trick for transforming the fundamental Hohenberg-Kohn variational principle to a system of differential equation for one-particle orbitals.

\subsection{Does TDCDFT/TDDFT have a predictive power?}

Despite conceptually the traditional formulation of TDCDFT (as well as TDDFT) looks clean and very similar to the static DFT, it may cause some confusions \cite{SchDre2007}. Let us assume that the stress tensor functional is known. The main goal of the theory is to calculate the current and the density at any instant $t$ by propagating the equations of motion (\ref{continuity1}) and (\ref{NS1}) starting from the initial time, provided the initial conditions for ${\bf j}$ and $n$ are given. Apparently the propagation of Eq.~(\ref{NS1}) is only possible if $\Pi_{\mu\nu}[{\bf j}]$ is a {\it retarded} functional of the current. In other words at a given instant $t$ the stress tensor should only depend on the currents ${\bf j}(t')$ at previous times $t'<t$. In particular, it should not contain local in time terms which depend on the time derivatives (of any order $r\ge 1$) of the current. However, both the original Runge-Gross proof \cite{RunGro1984}, and its generalization to TDCDFT by Ghosh and Dhara \cite{GhoDha1988} establish only the uniqueness of the maps ${\bf j}\mapsto{\bf A}$ and ${\bf j}\mapsto\Psi$ , but do not state anything about the properties of functionals ${\bf A}[{\bf j}]$ and $\Psi[{\bf j}]$. On the other hand if we consider the local momentum conservation law, Eq.~(\ref{NS1}), and interpret it as an equation which determines the external 4-potential for a given current, we immediately observe that the external force, and thus the external potential considered as functional of the current does indeed contain a local in time term proportional to $\partial_t{\bf j}$. Taking this fact naively one may conclude that the functionals $\Psi[{\bf j}]$ and $\Pi_{\mu\nu}[{\bf j}]$ should also contain such unwanted terms. Therefore the propagation of the basic equation Eq.~(\ref{NS1}) (or equivalently Eq.~(\ref{NS2})) becomes problematic. Hence it appears that TDCDFT/TDDFT does not have a real predictive power. 

In a recent paper Maitra et.~al. clearly demonstrated that the above conclusion is incorrect for the KS formulation of TDDFT \cite{MaiLeeBur2008}. In fact, this is obvious from the van~Leeuwen proof of the TDDFT mapping theorem \cite{vanLeeuwen1999}, and its generalization to TDCDFT by Vignale \cite{Vignale2004}. The construction proposed in Refs.~\cite{vanLeeuwen1999,Vignale2004} shows that the {\it difference} of potentials driving the dynamics two different systems with the same current (e.~g., interacting and noninteracting systems in the KS case) is uniquely recovered from the given density/current and has a proper causal structure. Unfortunately this methodology does not allow, at least directly, to make any definite statement about the retardation properties of the many-body wave function $\Psi[{\bf j}]$ as a functional of the current and, what is much more important, about the causal structure of the stress tensor functional entering the collective variable theory defined by Eqs.~(\ref{continuity2}) and (\ref{NS2}).  In Sec.~\ref{TDDefFT} we show how these questions are resolved within a new constructive reformulation of the theory based on TDDefFT. As a first step in this direction we discuss one more alternative form of the TDCDFT hydrodynamics.

\subsection{TDCDFT hydrodynamics in the Lagrangian form}

In general TDCDFT is a closed formalism that allows one to describe a convective motion of a quantum many-body system driven by an external field. Usually the convective motion is characterized by the density of particles $n({\bf x},t)$ and the density of current ${\bf j}({\bf x},t)$ or a velocity field ${\bf v}({\bf x},t)$. An alternative way to completely characterize the convective motion is commonly referred to as a Lagrangian description. Let us consider the system as a collection of infinitesimal fluid elements (so called ``materials''). Every fluid element can be uniquely labeled by a continuous variable $\bm\xi$ -- its position at the initial time $t=0$. The Lagrangian description can be viewed as tracking the motion of those infinitesimal elements of the fluid. In other words the convective motion of the system is characterized by a (continuous) set of trajectories ${\bf x}(\bm\xi,t)$, where the argument $\bm\xi$ indicates the starting point of the trajectory (the unique label of the element). Let us show that in the case of dynamics of a quantum many-body system the map ${\bf v}({\bf x},t)\mapsto{\bf x}(\bm\xi,t)$ is unique and invertible. 

For a given velocity ${\bf v}({\bf x},t)$ the Lagrangian trajectory  is solution to the following Cauchy problem
\begin{equation} 
\partial_t {\bf x}(\bm\xi,t) 
= {\bf v}({\bf x}(\bm\xi,t),t), \qquad {\bf x}(\bm\xi,0) = \bm\xi.
\label{traj}
\end{equation}
It is known (see, for example, Ref.~\cite{ODEbyArnold}) that Eq.~(\ref{traj}) has a unique solution ${\bf x}(\bm\xi,t)$ if the function ${\bf v}({\bf x},t)$ is continuous and satisfies the Lipschitz condition in spatial variables (i.~e. there exists a constant $L>0$, such that $\left\|{\bf v}({\bf x})-{\bf v}({\bf x}')\right\|<L\left\|{\bf x}-{\bf x}'\right\|$ for any ${\bf x}$ and ${\bf x}'$). Apparently the velocity field coming from a physical wave function does satisfy these requirements. Physically the Lipschitz condition prevents generation of folds and shock fronts which are clearly absent at the microscopic level for the Schr\"odinger dynamics. Hence from a given physical velocity field we uniquely construct a set of Lagrangian trajectories. Every trajectory ${\bf x}(\bm\xi,t)$ is uniquely determined by its initial point $\bm\xi$, which means that given the initial position $\bm\xi$ of a fluid element we can always find its coordinate ${\bf x}={\bf x}(\bm\xi,t)$ at any instant $t$, and, tracing the trajectory in a reverse order, from a given position ${\bf x}$ at time $t$ one uniquely recovers the initial point of the trajectory, $\bm\xi=\bm\xi({\bf x},t)$. In a more formal language, the map $\bm\xi\mapsto{\bf x}:{\bf x}={\bf x}(\bm\xi,t)$ is unique and invertible. This fact allows us to recover the Eulerian variables, i.~e., the velocity field and the density, from given Lagrangian trajectories
\begin{eqnarray}
\label{vfromxi}     
{\bf v}({\bf x},t) &=& \left[
\frac{\partial{\bf x}(\bm\xi,t)}{\partial t}
\right]_{\bm\xi = \bm\xi({\bf x},t)},\\
n({\bf x},t) &=& \left[
\frac{n_{0}(\bm\xi)}{\sqrt{g(\bm\xi,t)}}
\right]_{\bm\xi = \bm\xi({\bf x},t)},
\label{nfromxi} 
\end{eqnarray}
where $\bm\xi({\bf x},t)$ is the inverse of ${\bf x}(\bm\xi,t)$, $n_0({\bf x})$ is the initial density, and $\sqrt{g(\bm\xi,t)}=J(\bm\xi,t) = \det(\frac{\partial x^{\mu}}{\partial\xi^{\nu}})$ is the Jacobian of the transformation of coordinates ${\bf x}\to \bm\xi$. Equation (\ref{vfromxi}) is an obvious consequence of Eq.~(\ref{traj}), while Eq.~(\ref{nfromxi}) is a direct solution of the continuity equation (\ref{continuity2}). Therefore the function ${\bf x}(\bm\xi,t)$ indeed completely characterizes the convective motion of a system.

The basic equation in the Lagrangian description of collective dynamics is the equation of motion for a fluid element. This equation can be straightforwardly derived from the equation of motion for Eulerian velocity ${\bf v}({\bf x},t)$, Eq.~(\ref{NS2}), by making a transformation coordinates ${\bf x}\to \bm\xi$, i.~e., by considering the initial points $\bm\xi$ of Lagrangian trajectories as independent spatial coordinates. Under this transformation the convective derivative, $\partial_t + {\bf v}\partial_{\bf x}$ becomes simply $\partial_t$ so that the first term in Eq.~(\ref{NS2}) transforms to $m\ddot{\bf x}(\bm\xi,t)$, while the divergence of the stress tensor in the last term in Eq.~(\ref{NS2}) becomes a covariant divergence in the space with metrics $g_{\mu\nu}(\bm\xi,t)$ induced by the transformation from ${\bf x}$- to $\bm\xi$-coordinates. Hence after the transformation of coordinates we arrive at the following equation of motion for a fluid element
\begin{equation}
 \label{NSLagr}
m\ddot{x}^{\mu} - E_{\mu}({\bf x},t) - [\dot{\bf x}\times{\bf B}({\bf x},t)]_{\mu} 
+ \frac{\sqrt{g}}{n_0}\frac{\partial \xi^{\alpha}}{\partial x^{\mu}}
\nabla_{\nu}\widetilde{P}^{\nu}_{\alpha}[{\bf x}(\bm\xi,t)]=0,
\end{equation}
where $\widetilde{P}_{\mu\nu}(\bm\xi,t)$ is the original stress tensor $P_{\alpha\beta}({\bf x},t)$ transformed to the new coordinates according to the standard rules \cite{DubrovinI}
\begin{equation}
 \label{tildeP}
\widetilde{P}_{\mu\nu}(\bm\xi,t) =
\frac{\partial x^{\alpha}}{\partial \xi^{\mu}}\frac{\partial x^{\beta}}{\partial \xi^{\nu}} 
P_{\alpha\beta}({\bf x}(\bm\xi,t),t)
\end{equation}
The nabla-operator in Eq.~(\ref{NSLagr}) stands for a covariant divergence
\begin{equation} 
\nabla_{\nu}\widetilde{P}^{\nu}_{\mu} = \frac{1}{\sqrt{g}}
\partial_{\xi^{\nu}}\sqrt{g}\widetilde{P}^{\nu}_{\mu} 
- \frac{1}{2}\widetilde{P}^{\alpha\beta}\partial_{\xi^{\mu}}g_{\alpha\beta},
\label{nabla}
\end{equation}
and the metric tensor in the $\bm\xi$-space of ``initial positions'' is defined as follows
\begin{equation} 
g_{\mu\nu}(\bm\xi,t) = \frac{\partial x^{\alpha}}{\partial\xi^{\mu}}
\frac{\partial x^{\alpha}}{\partial\xi^{\mu}}; \quad
[g_{\mu\nu}]^{-1} = g^{\mu\nu} = \frac{\partial\xi^{\mu}}{\partial x^{\alpha}}
\frac{\partial\xi^{\nu}}{\partial x^{\alpha}}
\label{metric}
\end{equation}
The equation of motion (\ref{NSLagr}) has to be solved with the initial conditions ${\bf x}(\bm\xi,0)=\bm\xi$ and $\dot{\bf x}(\bm\xi,0)= {\bf v}_0(\bm\xi)$, where ${\bf v}_0({\bf x})$ is the initial velocity distribution calculated from the initial many-body wave function of Eq.~(\ref{InitialPsi}).

The first three terms in Eq.~(\ref{NSLagr}) correspond to a classical Newton equation for a particle moving in the external electromagnetic field, while the last, stress term takes care of all complicated quantum and many-body effects. Because of the uniqueness and invertibility of the map ${\bf v}({\bf x},t)\mapsto{\bf x}(\bm\xi,t)$ the transformed stress tensor  can be considered as a unique functional of the Lagrangian trajectories, $\widetilde{P}_{\mu\nu}=\widetilde{P}_{\mu\nu}[{\bf x}(\bm\xi,t)]$. Hence Eq.~(\ref{NSLagr}) is a closed equation of motion, which, at the formally exact level, completely determines the collective dynamics of the system. This is the basic equation of TDCDFT in the Lagrangian form.

From the first sight the representation of TDCDFT hydrodynamics in the Lagrangian form of Eq.~(\ref{NSLagr}) does not bring anything fundamentally new. This is indeed true if one follows a route outlined in this section: starting from the traditional formulation of the TDCDFT mapping theorem and via the Eulerian equation of motion for the density and the current. However, in the next section we will see that using the ideas of the Lagrangian description one can reformulate the whole theory, including all basic theorems, in a constructive way that also ends up with the equation of motion (\ref{NSLagr}), but provides us with a clear constrained search-like procedure for calculating the basic stress tensor functional. 

\section{Time-dependent deformation functional theory}
\label{TDDefFT}

\subsection{Many-body theory in a comoving reference frame}

Conceptually TDCDFT is a reduced theory aimed at describing only the convective motion of the system without a detailed knowledge of the full dynamics of all microscopic degrees of freedom. Therefore it looks natural to start the consruction of such a theory by separating the ``convective'' degrees of freedom at the very beginning, i.~e. at the level of the full microscopic many-body theory. The Lagrangian description is perfectly suited for this purpose. Indeed, in this formalism the convective dynamics is characterized by the motion of fluid elements. Therefore it can be easily separated from the microscopic dynamics of quantum particles by transforming the many-body theory to a local noninertial reference frame moving along the Lagrangian trajectories. 

At the formal level one proceeds as follows. Consider a reference frame defined by some (unspecified for the moment) velocity field ${\bf v}({\bf x},t)$ which is required to be continuous and Lipschitz in spatial variables. By solving the Cauchy problem of Eq.~(\ref{traj}) with the above velocity in the right hand side we get the local trajectories ${\bf x}(\bm\xi,t)$ of the frame. The transformation of the theory to the new reference frame corresponds to a transformation of coordinates ${\bf x}_j\to \bm\xi_j$, with ${\bf x}(\bm\xi,t)$ being the transformation function, i.~e. ${\bf x}_j={\bf x}(\bm\xi_j,t)$, in the many-body Schr\"odinger equation (\ref{SE}). It is convenient to define the many-body wave function $\widetilde{\Psi}({\bm\xi}_{1},\dots,{\bm\xi}_{N},t)$ in the new frame as follows \cite{TokatlyPRB2007}
\begin{equation}
\widetilde{\Psi}({\bm\xi}_{1},\dots,{\bm\xi}_{N},t)
= \prod_{j = 1}^{N}g^{\frac{1}{4}}({\bm\xi}_{j},t)
e^{-iS_{\text{cl}}({\bm\xi}_{j},t)}\Psi({\bf x}({\bm\xi}_{1},t),\dots,{\bf x}({\bm\xi}_{N},t),t),
\label{PsiLagr}
\end{equation}
where $S_{\text{cl}}({\bm\xi},t)$ is the classical action of a particle moving along the trajectory ${\bf x}(\bm\xi,t)$
\begin{equation}
\label{action}
S_{\text{cl}}({\bm\xi},t)= \int_0^t\left[ 
\frac{m}{2}(\dot{\bf x}(t))^{2} 
+ \dot{\bf x}(t){\bf A}({\bf x}(t),t) - U({\bf x}(t),t)\right].
\end{equation}
Equation (\ref{PsiLagr}) is a relatively straightforward generalization of the transformation to a homogeneously accelerated frame, which is used, for example, in the proofs of a harmonic potential theorem \cite{Dobson1994,Vignale1995a}. The exponential prefactor accounts for the phase acquired due to motion of the frame, while the factor $\prod_{j = 1}^{N}g^{\frac{1}{4}}({\bm\xi}_{j},t)$ is aimed at preserving the standard normalization of the wave function $\langle\widetilde{\Psi}|\widetilde{\Psi}\rangle=1$ after a non-volume-preserving transformation of coordinates.

Performing a transformation of coordinates, ${\bf x}_j\to \bm\xi_j$, in Eq.~(\ref{SE}), and using the definition (\ref{PsiLagr}) we obtain the many-body Schr\"odinger equation in the frame moving with some velocity ${\bf v}({\bf x},t)$
\begin{equation}
 \label{SELagr}
i\partial_t\widetilde{\Psi}({\bm\xi}_{1},\dots,{\bm\xi}_{N},t) =
\widetilde{H}[g_{ij},\bm{\mathcal A}]\widetilde{\Psi}({\bm\xi}_{1},\dots,{\bm\xi}_{N},t)
\end{equation}
The Hamiltonian in the new frame takes the form
\begin{equation}
 \label{HLagr}
\widetilde{H}[g_{ij},\bm{\mathcal A}] = \sum_{j = 1}^{N}g^{-\frac{1}{4}}_{j}
\hat{K}_{j,\mu}\frac{\sqrt{g_{j}}g^{\mu\nu}_{j}}{2m}
\hat{K}_{j,\nu}g^{-\frac{1}{4}}_{j}
+\frac{1}{2}\sum_{k\ne j}V(l_{\bm\xi_{k}\bm\xi_{j}})
\end{equation}
where $\hat{K}_{j,\mu}=-i\partial_{\xi^{\mu}_{j}}
- {\cal A}_{\mu}(\bm\xi_{j},t)$, $g^{\mu\nu}_{j}=g^{\mu\nu}(\bm\xi_{j},t)$, and $l_{\bm\xi_{k}\bm\xi_{j}}$ is the distance between $j$th and $k$th particles in the moving frame (the length of geodesic connecting points $\bm\xi_{j}$ and $\bm\xi_{k}$ in the space with metric $g_{\mu\nu}$). An effective vector potential $\bm{\mathcal A}(\bm\xi,t)$ entering the Hamiltonian is given by the following equation
\begin{equation}
 \label{Aeff}
{\mathcal A}_{\mu} = \frac{\partial x^{\nu}}{\partial\xi^{\mu}}\dot{x}^{\nu} + 
\frac{\partial x^{\nu}}{\partial\xi^{\mu}}A_{\nu}({\bf x},t) - \partial_{\xi^{\mu}}S_{\text{cl}}({\bm\xi},t).
\end{equation}
Physically this effective vector potential describes a combined action of the external and inertial forces in a local noninertial frame. Since we defined the moving frame in such a way that at $t=0$ it coinsides with the laboratory one (this formally follows from the condition ${\bf x}(\bm\xi,0) = \bm\xi$) the initial condition to the transformed Schr\"odinger equation (\ref{SELagr}) remains unchanged:
\begin{equation}
\widetilde{\Psi}({\bm\xi}_{1},\dots,{\bm\xi}_{N},0)= \Psi_0({\bm\xi}_{1},\dots,{\bm\xi}_{N}).
 \label{InitialPsiLagr}
\end{equation}
Equations (\ref{NSLagr})-(\ref{InitialPsiLagr}) completely determine dynamics of the quantum $N$-particle system in an arbitrary local noninertial frame.

Since our aim was to reformulate the theory in a particular frame moving with a physical flow (this frame is called comoving, or Lagrangian) we need to impose an additional local condition to specify the required frame. By definition in the comoving frame the current density is zero everywhere and at all times, while the density of particles stays stationary and equal to the initial density distribution $n_0(\bm\xi)$. Hence the most natural frame-fixing condition is the requirement of zero transformed current density $\widetilde{\bf j}(\bm\xi,t)=0$. Explicitly this condition reads
\begin{equation}
 \label{gauge}
\bm{\mathcal A}(\bm\xi,t) = \frac{i}{2n_0(\bm\xi)}
\lim_{{\bm\xi}'\to{\bm\xi}}(\partial_{\bm\xi} -\partial_{\bm\xi'})\widetilde{\rho}({\bm\xi},{\bm\xi}',t)
\end{equation}
where $\widetilde{\rho}({\bm\xi},{\bm\xi}',t)$ is the one particle reduced density matrix calculated from the transformed wave function
\begin{equation}
 \label{rhoLagr}
\rho({\bf x},{\bf x}',t) = N \int \prod_{j =2}^{N}d{\bm\xi}_{j}
\Psi^*({\bm\xi},{\bm\xi}_{2},\dots,{\bm\xi}_{N},t)\Psi({\bm\xi}',{\bm\xi}_{2},\dots,{\bm\xi}_{N},t).
\end{equation}
The frame-fixing condition (\ref{gauge}) simply states that in the comoving frame a ``paramagnetic'' current (the right hand side of (\ref{gauge})) is exactly cancelled by the ``diamagnetic'' contribution (the left hand side of (\ref{gauge})).

The Schr\"odinger equation (\ref{SELagr}), the definition of the effective vector potential (\ref{Aeff}), and the zero current condition (\ref{gauge}) constitute a closed system of equations that determine dynamics of the many-body system in the comoving reference frame. In principle one can eliminate the effective vector potential from this system by substituting $\bm{\mathcal A}(\bm\xi,t)$ from Eq.~(\ref{gauge}) into Eqs.~(\ref{Aeff}) and (\ref{SELagr}). As a result we will get a system of two first order (in time) differential equation for two functions -- ${\bf x}(\bm\xi,t)$ which describes the convective motion on the system, and $\widetilde{\Psi}({\bm\xi}_{1},\dots,{\bm\xi}_{N},t)$ describing the rest of microscopic degrees of freedom in the frame moving with the convective flow. The equations have to be solved with the initial conditions Eq.~(\ref{InitialPsiLagr}) for $\widetilde{\Psi}(t)$ and  ${\bf x}(\bm\xi,0) = \bm\xi$ for the Lagrangian trajectory. The system of Eqs.~(\ref{SELagr}), (\ref{Aeff}) and (\ref{gauge}) is equivalent to the original linear Schr\"odinger equation (\ref{SE}). Therefore we can guarantee that there exists a unique solution of the corresponding initial value problem.

\subsection{A ``constrained search'' formulation of TDDefFT}

Let us discuss possible procedures for solving the system of Eqs.~(\ref{SELagr}), (\ref{Aeff}) and (\ref{gauge}), but, before doing that, it makes a sense to rewrite it in a more physical and clear form. 

First of all, we note that because Eq.~(\ref{Aeff}) contains the classical action $S_{\text{cl}}({\bm\xi},t)$ of Eq.~(\ref{action}), it is nonlocal in time. It is convenient to remove this nonlocality by differentiating Eq.~(\ref{Aeff}) with respect to $t$. Taking the derivative we reduce Eq.~(\ref{Aeff}) to the following form
\begin{equation}
 \label{traj1}
m\ddot{x}^{\mu} = E_{\mu}({\bf x},t) + [\dot{\bf x}\times{\bf B}({\bf x},t)]_{\mu} 
+ \frac{\partial \xi^{\nu}}{\partial x^{\mu}}
\partial_{t}{\cal A}_{\nu}, 
\end{equation}
which is exactly the classical Newtonian equation for a particle moving in the external electromagnetic field and, in addition, in the ``electric'' field. generated by the effective vector potential. 

Comparing Eq.~(\ref{traj1}) with Eq.~(\ref{NSLagr}) we immediately observe that these two equation would be identical if the time derivative of the effective vector potential would equal to the covariant divergence of the stress tensor. To see that this is indeed the case we consider the local force balance equation that follows from the many-body Schr\"odinger equation (\ref{SELagr}) in the local noninertial frame. Apparently this equation is of the form of Eq.~(\ref{NS1}), but with the usual divergence of the momentum flow tensor replaced by covariant divergence of the stress tensor in the $\bm\xi$-space. Namely,
\begin{equation}
 \label{NSg}
\partial_{t}\widetilde{j}_{\mu} -  \widetilde{j}^{\nu}
(\partial_{\xi^{\nu}}{\cal A}_{\mu} - \partial_{\xi^{\mu}}{\cal A}_{\nu})
+ \widetilde{n}\partial_{t}{\cal A}_{\mu}
+ \sqrt{g}\nabla_{\nu}\widetilde{P}^{\nu}_{\mu} = 0.
\end{equation}
where the stress tensor in the space with metric $g_{\mu\nu}$, can be conveniently determined from the following universal formula \cite{TokatlyPRB2005a,TokatlyPRB2007,RogRap2002}
\begin{equation}
 \label{StressLagr}
\widetilde{P}_{\mu\nu}(\bm\xi,t) = \frac{2}{\sqrt{g}}\langle\widetilde{\Psi}|
\frac{\delta\widetilde{H}[g_{\alpha\beta},
\bm{\mathcal{A}}]}{\delta g^{\mu\nu}(\bm\xi,t)}|\widetilde{\Psi}\rangle 
\equiv\langle\widetilde{\Psi}|\hat{\widetilde{P}}_{\mu\nu}[g_{\alpha\beta},\bm{\mathcal{A}}]|\widetilde{\Psi}\rangle 
\end{equation}
with the Hamiltonian defined by Eq.~(\ref{HLagr}). An explicit form of the stress tensor operator $\hat{\widetilde{P}}_{\mu\nu}$ entering Eq.~(\ref{StressLagr}) can be found, for example, in Ref.~\cite{TokatlyPRB2005a}. At the moment the important point is only that $\hat{\widetilde{P}}_{\mu\nu}[g_{\alpha\beta},\bm{\mathcal{A}}]$ is an explicitly known and local in time functional of the metric tensor and the effective vector potential. Since in the comoving frame the transformed current density $\widetilde{\bf j}$ is zero, only the last two terms survive in Eq.~(\ref{NSg}). Therefore in our frame of interest the force balance equation reduces to the following identity
\begin{equation}
 \label{balance}
 \partial_{t}{\cal A}_{\mu} = - \frac{\sqrt{g}}{n_0}\nabla_{\nu}\widetilde{P}^{\nu}_{\mu}
\end{equation}
Inserting this identity into Eq.~(\ref{traj1}) we exactly recover the basic equation of hydrodynamics in the Lagrangian description, Eq.~(\ref{NSLagr}). The important progress is that the stress tensor in this equation is now defined entirely in terms of the variables entering the many-body problem in the comoving frame. Hence we have transformed Eq.~(\ref{Aeff}) to a transparent physical form and demonstrated that it is indeed the correct equation of motion for the fluid elements.

Using the results of Eqs.~(\ref{traj1}) and (\ref{balance}) we can write down the following final system of equations describing the dynamics of the full many-body system
\begin{eqnarray}
 \label{SELfin}
&&i\partial_t\widetilde{\Psi}({\bm\xi}_{1},\dots,{\bm\xi}_{N},t) =
\widetilde{H}[g_{\mu\nu},\bm{\mathcal A}]\widetilde{\Psi}({\bm\xi}_{1},\dots,{\bm\xi}_{N},t)\\
\label{gaugefin}
&&\bm{\mathcal A}(\bm\xi,t) = \frac{i}{2n_0(\bm\xi)}
\lim_{{\bm\xi}'\to{\bm\xi}}(\partial_{\bm\xi} -\partial_{\bm\xi'})\widetilde{\rho}({\bm\xi},{\bm\xi}',t)\\
\label{NSLfin}
&&m\ddot{x}^{\mu} = E_{\mu}({\bf x},t) + [\dot{\bf x}\times{\bf B}({\bf x},t)]_{\mu} 
- \frac{\sqrt{g}}{n_0}\frac{\partial \xi^{\alpha}}{\partial x^{\mu}}
\nabla_{\nu}\widetilde{P}^{\nu}_{\alpha},
\end{eqnarray}
where the Hamiltonian $\widetilde{H}[g_{\mu\nu},\bm{\mathcal A}]$, the reduced density matrix $\widetilde{\rho}[\widetilde{\Psi}]({\bm\xi},{\bm\xi}',t)$, and the stress tensor $\widetilde{P}_{\mu\nu}[g_{\alpha\beta},\bm{\mathcal A},\widetilde{\Psi}]$ are defined after Eqs.~(\ref{HLagr}), (\ref{rhoLagr}), and (\ref{StressLagr}), respectively. The metric tensor $g_{\mu\nu}(\bm\xi,t)$ entering Eqs.~(\ref{SELfin}) and (\ref{NSLfin}) is related to the Lagrangian trajectory via Eq.~(\ref{metric}). 

It is important to stress out that the system of Eqs.~(\ref{SELfin})-(\ref{NSLfin}) is mathematically equivalent to the original many-body Schr\"odinger equation (\ref{SE}). In fact, everything we did to derive Eqs.~(\ref{SELfin})-(\ref{NSLfin}) from Eq.~(\ref{SE}) was an identical change of variables aimed at separating the convective and the ``relative'' motions of quantum particles. However after this identical transformation the structure of the many-body theory becomes quite remarkable. The key observation is that the physical external fields enter only the equation of motion for the fluid elements, Eq.~(\ref{NSLfin}), while the many-body dynamics, which is governed by Eqs.~(\ref{SELfin}) and (\ref{gaugefin}), depends only on the fundamental geometric characteristic of the comoving frame -- the metric tensor $g_{\mu\nu}(\bm\xi,t)$. Formally Eqs.~(\ref{SELfin}) and (\ref{gaugefin}) describe the dynamics of $N$ quantum particles driven by a given time-dependent metric and constrained by the requirement of zero current density. This constrained many-body problem can be equivalently represented in a form of a Dirac-Frenkel variational principle with the folowing action functional
\begin{equation}
 \label{variat}
S[\widetilde{\Psi},\bm{\mathcal A}] = \int_0^T dt\left(i\langle\widetilde{\Psi}^*|\partial_t |\widetilde{\Psi}\rangle
- \langle\widetilde{\Psi}^*|\widetilde{H}[g_{\mu\nu},\bm{\mathcal A}]|\widetilde{\Psi}\rangle\right).
\end{equation}
Indeed, the conditions for the extremum of this action,
\begin{equation}
 \nonumber
\frac{\delta S[\widetilde{\Psi},\bm{\mathcal A}]}{\delta\widetilde{\Psi}^*}=0, \qquad
\frac{\delta S[\widetilde{\Psi},\bm{\mathcal A}]}{\delta\bm{\mathcal A}}=0,
\end{equation}
are identical to Eqs.~(\ref{SELfin}) and (\ref{gaugefin}), respectively. From the variational formulation we clearly see the the effective vector potential is simply a Lagrange multiplier that ensures the zero current constraint.

The formulation of the many-body part of the problem in a form of the variational principle can be viewed as a time-dependent analog of the Levi-Lieb constrained search formulation of the static DFT \cite{Levy1979,Levy1982,Lieb1983}(see also \cite{DreizlerGross1990}). By finding an extremizer of the functional (\ref{variat}), or, equivalently,  by solving the constrained many-body problem of  Eqs.~(\ref{SELfin}) and (\ref{gaugefin}) for a given metric of the form (\ref{metric}) we get the wave function $\widetilde{\Psi}$ and the Lagrange multiplier $\bm{\mathcal A}$ as universal functionals of the metric tensor: $\widetilde{\Psi}=\widetilde{\Psi}[g_{\mu\nu}]$ and $\bm{\mathcal A}=\bm{\mathcal A}[g_{\mu\nu}]$. Substitution of these functionals in to Eq.~(\ref{StressLagr}) gives the universal stress tensor functional $\widetilde{P}_{\mu\nu}[g_{\alpha\beta}]$. Thus Eq.~(\ref{NSLfin}) becomes a closed equation of motion for fluid elements, which determines the Lagrangian trajectories of the system. As all basic quantities are functionals of the metric tensor, which physically is nothing but the Green's deformation tensor of the classical elasticity theory \cite{PhysAc}, it is natural to call this approach the time-dependent deformation functional theory (TDDeFT).

The direct solution of Eq.~(\ref{NSLfin}) with a known functional $\widetilde{P}_{\mu\nu}[g_{\alpha\beta}](\bm\xi,t)$ gives the description of the convective motion in terms of the Lagrangian picture. Alternatively we can transform $\widetilde{P}_{\mu\nu}[g_{\alpha\beta}](\bm\xi,t)$ to the laboratory frame to recover the tensor ${P}_{\mu\nu}[{\bf v}]({\bf x},t)$,
\begin{equation}
 \label{PfromTildeP}
{P}_{\mu\nu}[{\bf v}]({\bf x},t) =
\frac{\partial \xi^{\alpha}}{\partial x^{\mu}}\frac{\partial \xi^{\beta}}{\partial x^{\nu}} 
\widetilde{P}_{\alpha\beta}[g_{\alpha\beta}(\bm\xi({\bf x},t),t)](\bm\xi({\bf x},t),t),
\end{equation}
which can be used either in the hydrodynamic formulation of Eqs.~(\ref{continuity1})-(\ref{NS1}) or to calculate the xc potentials for the KS formulation of TDCDFT, Eqs.~(\ref{KS})-(\ref{Axc}). Finally, since in the laboratory frame the function $\bm\xi({\bf x},t)$ (the initial point of the trajectory that arrives to ${\bf x}$ at the time $t$) can be found from the equation
\begin{equation}
 \label{xi}
[\partial_t + {\bf v}({\bf x},t)\partial_{\bf x}]\bm\xi({\bf x},t)=0, \quad \bm\xi({\bf x},0)={\bf x}
\end{equation}
the stress tensor determind by Eq.~(\ref{PfromTildeP}) is indeed a universal functional of the Eulerian velocity ${\bf v}({\bf x},t)$.

Therefore we recovered the full formal structure of the traditional TDCDFT, but at the fundamentally new level of understanding. First of all, in TDDefFT formalism a closed theory of convective motion appears from a regular and conceptually clean procedure: it is simply a first universal step in solving the many-body problem in the comoving frame. Hence now we clearly understand where the universal functional entering the theory come from and why they are universal. Also, the causal structure of all functionals becomes absolutely transparent. Indeed, both the constraint of Eq.~(\ref{gaugefin}) and the right hand side in Eq.~(\ref{SELfin}) are local in time. Therefore the solution of the constrained problem of Eqs.~(\ref{SELfin}) and (\ref{gaugefin}), provided it exists, defines the wave function $\widetilde{\Psi}$ as a strictly retarded functional of the metric tensor. Hence both $\widetilde{P}_{\mu\nu}[g_{\alpha\beta}]$ and ${P}_{\mu\nu}[{\bf v}]$ depend, respectively, on $g_{\alpha\beta}$ and $g_{\alpha\beta}$ in a strictly retarded manner. Thus the new formulation completely removes all doubts concerning the predictable power of both hydrodynamic and KS formulation of TDCDFT.

\subsection{Basic uniqueness and existence theorems of TDDefFT}

The traditional formulation of TDDFT/TDCDFT is based on two key mathematical statements: (i) the uniqueness and invertibility of the mapping from external potentials to the density/current, and (ii) the v-representability of the density/current. At the present stage of the theory we have proofs \cite{RunGro1984,vanLeeuwen1999,GhoDha1988,Vignale2004} of the mapping theorem for Taylor expandable potentials, while the v-representability problem, strictly speaking, remains open. To prove the v-representability within techniques used in the available proofs of the mapping theorems on has to demonstrate that the uniquely constructed power series for potentials do converge, which has not been done up to now.

Within the new formulation based on TDDefFT there is no question of mapping as there is no external potential in the universal many-body problem in the comoving frame. Instead the two above mathematical issues reappear in a form of the uniqueness and existence of a solution to nonlinear system of Eqs.~(\ref{SELfin}), (\ref{gaugefin}). The uniqueness is a reminiscent of the potential-to-current mapping, while the problem of the existence of a solution is equivalent to the problem of interacting v-representability. Equations (\ref{SELfin}) and (\ref{gaugefin}) correspond to interacting system, but the same set of questions can be asked for a system of $N$ noninteracting particles. In the later case the dynamics is characterized by $N$ one particle orbitals $\varphi_j(\bm\xi,t)$, $j=1\dots N$, and the universal constrained many-body problem simplifies as follows
\begin{eqnarray}
 \label{SELnonint}
&&i\partial_t\varphi_j = g^{-\frac{1}{4}}
(i\partial_{\xi^{\mu}} + {\cal A}_{\mu})\frac{\sqrt{g}g^{\mu\nu}}{2m}
(i\partial_{\xi^{\nu}} + {\cal A}_{\nu})g^{-\frac{1}{4}}\varphi_j, \\
\label{gaugenonint}
&& \bm{\mathcal A} = \frac{-i}{2n_0}\sum_{j=1}^{N}
(\varphi^*_j\partial_{\bm\xi}\varphi_j - \varphi_j\partial_{\bm\xi}\varphi^*_j)
\end{eqnarray}
where $n_0(\bm\xi)=\sum_{j=1}^{N}|\varphi_j(\bm\xi,0)|^2=\sum_{j=1}^{N}|\varphi_j(\bm\xi,t)|^2$ is the density of particles which by construction is independent of time. The problem of existence of a solution to the system of Eqs.~(\ref{SELnonint}), (\ref{gaugenonint}) is similar to the well known problem of the noninteracting v-representability in the traditional formulation of the TDCDFT. Note that for the practical purpose of TDDefFT/TDCDFT it is not necessarily to prove the well-posedness of the systems Eqs.~(\ref{SELfin}), (\ref{gaugefin}) and Eqs.~(\ref{SELnonint}), (\ref{gaugenonint}) for any metric. It it enough to consider the metric tensors of the form (\ref{metric}), i.~e. the metrics which are generated by an invertible transformation coordinates and correspond to a flat space.

The key advantage the new formulation of old issues is that now the basic questions underlying TDCDFT are posed in the standard form common in mathematical physics. It is quite likely that the uniqueness and the existence theorems for nonlinear systems of Eqs.~(\ref{SELfin}), (\ref{gaugefin}) and Eqs.~(\ref{SELnonint}), (\ref{gaugenonint}) can be proved using standard methods of the functional analysis and the theory of differential equations (see, e.~g., \cite{Segal1981} and Chapter~X in Ref.~\cite{ReedSimonII}), which have been successfully applied to the analysis of the time-dependent Hartree and Hartree-Fock systems \cite{ChaGla1975,BovPraFan1976,Chadam1976}. Structurally the nonlinear systems of Eqs.~(\ref{SELfin}), (\ref{gaugefin}) and Eqs.~(\ref{SELnonint}), (\ref{gaugenonint}) are of the type of a time-dependent Schr\"odinger equation with a special cubic nonlinearity, which, to the best of my knowledge, have not been considered before. An encouraging observation is that, in contrast to most known systems nonlinear Schr\"odinger equations, our equations  are exactly integrable in the simple case of $N=1$ (when the systems of Eqs.~(\ref{SELfin}), (\ref{gaugefin}) and Eqs.~(\ref{SELnonint}), (\ref{gaugenonint}) coinside) \cite{TokatlyPRB2007}. The explicit solution solution for $N=1$ is
\begin{equation}
\widetilde{\Psi}(\bm\xi,t) = \sqrt{n_{0}(\bm\xi)}
e^{i\varphi(\bm\xi,t)}, \quad
{\mathcal A}_{\mu}(\bm\xi,t) = \partial_{\xi^{\mu}}\varphi(\bm\xi,t)
\label{PsiN1}
\end{equation}
where
\begin{equation}
\varphi = \varphi_{0}(\bm\xi) +
\frac{1}{2m\sqrt{n_{0}}}\int_{0}^{t}\Big[g^{-\frac{1}{4}}\partial_{\xi^{\mu}}
\sqrt{g}g^{\mu\nu}\partial_{\xi^{\nu}}g^{-\frac{1}{4}}\sqrt{n_{0}}
\Big]dt'
\label{phaseN1}
\end{equation}
and $\varphi_{0}(\bm\xi)$ is a phase of the initial state. The corresponding stress tensor functional, Eq.~(\ref{StressLagr}), takes the form
\begin{equation}
\widetilde{P}_{\mu\nu}[g_{\alpha\beta}] = \frac{1}{m}\Big[
\big(\partial_{\xi^{\mu}}g^{-\frac{1}{4}}\sqrt{n_{0}}\big)
\big(\partial_{\xi^{\nu}}g^{-\frac{1}{4}}\sqrt{n_{0}}\big)
- \frac{g_{\mu\nu}}{4\sqrt{g}}\partial_{\xi^{\alpha}}\sqrt{g}g^{\alpha\beta}
\partial_{\xi^{\beta}}\frac{n_{0}}{\sqrt{g}}
\Big]
\label{StressN1}
\end{equation}
Equations (\ref{PsiN1}) and (\ref{phaseN1}) clearly show that the wave function and the effective vector potential are retarded functionals of the metric tensor. The stress tensor for the one particle case turns out to be absolutely local in time. Adding more particles should make the time dependence nonlocal (though necessarily retarded), but hopefully it will not spoil the existence and uniqueness of a solution.

Apparently, a Taylor expansion-based proof of uniqueness also works in the new setting, and actually becomes almost trivial. To show this we first substitute the constraint (\ref{gaugefin}) into Eq.~(\ref{SELfin}) to get a closed nonlinear evolution equation of the form
\begin{equation}
 \label{NlSE}
i\partial_t\widetilde{\Psi} = \widetilde{H}[g_{\mu\nu},\widetilde{\Psi}]\widetilde{\Psi}
\end{equation}
with a local in time nonlinear Hamiltonian. Let us assume that the metric tensor $g_{\mu\nu}$ is v-representable. In other words, it is given by Eq.~(\ref{metric}) with a Lagrangian trajectory that corresponds to a convective motion in some external potential. This is equivalent to the assumption that the solution to Eq.~(\ref{NlSE}) does exists. If in addition we assume that the metric possesses a covergent Taylor expansion in time, we can represent both $g_{\mu\nu}(t)$ and $\widetilde{\Psi}(t)$ in a form of power series
$$
g_{\mu\nu}(t) = \delta_{ij} + \sum_{k=1}^{\infty}\frac{g_{\mu\nu}^{(k)}}{k!}t^{k}, \quad
\widetilde{\Psi}(t) = \Psi_{0} + \sum_{k=1}^{\infty}\frac{\widetilde{\Psi}^{(k)}}{k!}t^{k}, 
$$
and insert them into Eq.~(\ref{NlSE}). Since the right hand side of Eq.~(\ref{NlSE}) is local in time all the coefficients $\widetilde{\Psi}^{(k)}$ with any $k$ are trivially expressible by recursion in terms of $\Psi_{0}$ and $g_{\mu\nu}^{(l)}$ with $l<k$. The later property is another direct manifestation of the retarded character of the functional $\widetilde{\Psi}[g_{\mu\nu}](t)$. Thus we have proved that for a given v-representable metric the solution is unique and has correct causal properties required for TDDefFT/TDCDFT to be a predictable theory in any formulation. 

If we relax the v-representability assumption we would have to prove that the uniquely constructed power series for $\widetilde{\Psi}(t)$ is convergent, which seems to be technically a very hard task. Most likely, early or later, the existence theorem will be proved by more advanced mathematical methods, applications of which becomes possible within the present reformulation of the theory. It is also quite likely that  the physically unmotivated requirement of the Taylor expandability will be removed, as it can be done in the linear responce regime of TDDFT \cite{vanLeeuwenTDDFT2006}.

\section{Conclusion}

In this work we proved the equivalence of two approaches to the convective dynamics of a general quantum many-body system and reconstructed TDCDFT on the new grounds. The most important outcome is the possibility to formulate TDCDFT within a conceptually clean and straightforward two step procedure which resembles the constrained search formulation of the static DFT. On the first step one solves a constrained time-dependent many-body problem to find a stress tensor as a universal functional of the deformation tensor. On the second step we use this functional to calculate the evolution of the current and the density for a given configuration of external fields. In this formulation the the vector potential-current density mapping theorem, and the v-representability problem are restated as the uniqueness and the existence theorems for a solution of a certain time-dependent nonlinear Schr\"odinger equation. Hence the fundamental questions of TDDFT appear in a standard setting of mathematical physics.
To my knowledge a very special type of nonlinearity appearing in TDDefFT has not been studied before. One of the hopes is that the new restatement of the basic theorems combined with a growing practical popularity of TDDFT/TDCDFT will attract attention of mathematicians and mathematical physicists to the formal foundations of the theory. I also believe that a clear physical picture behind the present approach should stimulate further developments of the theory within more physically oriented part of the community. In any case TDDFT and TDCDFT is still an exciting, active and promising area of research. Clearly, there is still much to be done both to put the theory on really firm mathematical grounds, and to develop practically working approximate functionalfor numerous applications.

This work was suported by the Ikerbasque Foundation.

\clearpage

\end{document}